\documentstyle[12pt]{article}
\newcommand{\PP}{{\sf I \hspace{-0.12em} P}}
\newcommand{\pmb}[1]{%
        \setbox0=\hbox{#1}%
        \kern-.02em\copy0\kern-\wd0
        \kern+.04em\copy0\kern-\wd0
        \kern-.02em\raise.0217em\box0}
\newcommand{\vek}[1]{
        \mathchoice{\mbox{\boldmath$#1$}}%
        {\mbox{\boldmath$#1$}}%
        {\pmb{$\scriptstyle#1$}}% 
        {\pmb{$\scriptscriptstyle#1$}}}
\newcommand{\lsim}{%less than or approx. symbol
 \mathrel{\setbox0=\hbox{$<$}\raise0.6ex\copy0\kern-\wd0
 \lower0.65ex\hbox{$\sim$}}}
\newcommand{\gsim}{%less than or approx. symbol
 \mathrel{\setbox0=\hbox{$>$}\raise0.6ex\copy0\kern-\wd0
 \lower0.65ex\hbox{$\sim$}}}

\begin{document}

\begin{titlepage}
\vspace*{-2cm}
\begin{flushright}
\bf 
TUM/T39-96-27
\end{flushright}

\bigskip 

\begin{center}
{\large \bf Diffractive Phenomena and Shadowing in 
            Deep-Inelastic Scattering$^{*)}$}

\vspace{2.cm}

{\large G. Piller, G. Niesler and W. Weise}  

\bigskip

Physik Department, Technische
Universit\"at M\"unchen
\\ D-85747 Garching, Germany\\ 

\vspace{5.cm}

{\bf Abstract}

\medskip
\begin{minipage}{15cm}
Shadowing effects in deep-inelastic lepton-nucleus  
scattering probe  the mass spectrum of diffractive 
leptoproduction from individual nucleons. 
We explore this relationship using current experimental 
information on both processes. 
In recent data from the NMC and E665 collaboration, taken at 
small $x\ll 0.1$ and $Q^2\lsim 1\,GeV^2$, 
shadowing is dominated by the diffractive excitation 
and coherent interaction of low mass vector mesons. 
If shadowing is explored at  
small $x\ll 0.1$ but large $Q^2\gg 1\,GeV^2$ as discussed 
at HERA, the situation is different.   
Here dominant contributions come  from the coherent 
interaction of diffractively produced heavy mass states. 
Furthermore we observe that the energy dependence of shadowing 
is directly related to the mass dependence of the 
diffractive production cross section for free nucleon targets.
\end{minipage}

\end{center}

\vspace*{1.cm}

%PACS numbers: 13.60.-r, 13.60.Le, 12.40.Vv

\vspace*{4.cm}

\noindent $^{*}$) Work supported  in part by BMBF.

\vfill
\newpage
\end{titlepage}

\section{Introduction}

In recent years diffractive photo- and leptoproduction of hadrons 
has become a field of growing interest. 
The diffractive excitation of heavy hadronic states with 
invariant masses $M_X\lsim 5 \,GeV$  has been investigated 
for photoproduction at FNAL \cite{Chapin85}. 
Recent data from HERA explore the kinematic region 
$8\,GeV^2<Q^2 < 100\,GeV^2$ and $M_X \lsim  15 \,GeV$ 
\cite{H195_diff,H196_diff,ZEUS95_diff,ZEUS96_diff,ICHEP_96}.

While diffractive photo- and leptoproduction of hadrons from 
nucleons is an interesting subject all by itself, 
it also plays a major role in the shadowing phenomena observed in 
high energy (virtual) photon-nucleus interactions. 
Shadowing in deep-inelastic lepton scattering is  important at 
small values of the Bjorken scaling variable $x=Q^2/2p\cdot q$, 
where $p^{\mu}$ and $q^{\mu} = (\nu,\vek q)$ are the nucleon and photon 
four-momenta, respectively, and $Q^2 = - q^2$. 
Here plenty of new data have become available from 
high precision experiments at FNAL (E665)  \cite{E665d,E665} 
and CERN (NMC) \cite{NMC}. 
Because of the fixed target nature of these experiments 
the average available momentum transfer is small, 
$Q^2 \lsim 1\,GeV^2$ at $x\ll 0.1$. 
Currently options to accelerate ions at HERA are under discussion. 
This would allow to investigate shadowing effects at $x\ll 0.1$ 
also at large $Q^2\gg 1\,GeV^2$ \cite{Nuclei_HERA}. 

In this paper we systematically explore nuclear shadowing in 
different kinematic 
regimes using the available experimental information on 
diffractive production processes from nucleons. 
Since the connection between diffraction and shadowing is 
most evident for  processes involving a pair of target nucleons, 
we will focus on shadowing effects in deuterium.    
We find that in the kinematic regime of the E665 and NMC measurements 
shadowing is mainly due to the diffractive excitation and 
multiple scattering of vector mesons.
This would be different at HERA, where the diffractive production 
of heavy mass states with $M_X^2 \sim Q^2$ would play a dominant role. 

Nuclear shadowing in deep-inelastic scattering can also 
reveal details on diffractive processes from individual nucleons. 
For example, the energy dependence of nuclear shadowing is 
directly related to the dependence of the 
diffractive production cross section on $M_X$, the mass of the 
final hadronic state. 

This paper is organized as follows: 
the available data on diffractive photo- and leptoproduction 
of hadrons are briefly summarized and discussed in Sec.2.
In Sec.3 we review the relationship between shadowing and diffraction. 
The diffractive production cross sections from  Sec.2 
are then used to investigate shadowing for deuterium. 
Conclusions follow  in Sec.4.

\section{Diffractive production of hadrons from free nucleons}
 
Consider the diffractive production of hadrons  
in the interaction  of high energy real or 
virtual photons with free nucleons,  
$ \gamma^{*} + N \rightarrow X + N$.
As in diffractive hadron-hadron collisions 
such processes are important at small transfered momenta. 
The corresponding cross sections  drop exponentially 
with $t= (p - p')^2 = k^2 \approx - \vek k^2$, 
where $p$ and $p'$ are the momenta of the initial and final nucleon. 
For our considerations we choose the laboratory frame 
where the target is at rest, $p^{\mu} = (M,{\bf 0})$, and fix  
the $z$-axis to be parallel to the 
photon momentum: $q^{\mu} = (\nu,{\bf 0}_{\perp},\sqrt{Q^2+\nu^2})$. 
The production of a hadronic final state $X$ with 
an invariant mass $M_X$ requires a minimal 
momentum transfer:
\begin{eqnarray} \label{eq:long} 
k_{z,min} &\simeq& 
\frac{Q^2 + M_X^2}{2 \nu} 
= M x \left( 1 + \frac{M_X^2}{Q^2} \right).
\end{eqnarray}
The diffractive excitation, with $t$ required to be small, 
of heavy hadronic states is therefore possible  only at sufficiently 
large photon energies $\nu$ or, equivalently, 
at small values of the Bjorken variable $x$.

In diffractive leptoproduction it is common to introduce 
the variable 
\begin{equation}
x_{\PP} =  \frac{(p-p')\cdot q}{p\cdot q} = 
\frac{Q^2 + M_X^2 - t}{Q^2 + W^2  - M^2}
\approx  
\frac{Q^2 + M_X^2}{Q^2 + W^2}, 
\end{equation}
with $W^2 = (p+q)^2 \approx 2M\nu - Q^2$. 
The notation $x_{\PP}$ is a reminder of pomeron phenomenology. 
In terms of this variable, 
\begin{equation}
k_{z,min} \approx M x_{\PP}.
\end{equation}

Our aim here is to investigate the relation  between  diffractive 
(virtual) photoproduction of hadrons from nucleons and shadowing corrections 
in deep-inelastic lepton-nucleus scattering. 
For this purpose we need to consider the forward 
diffractive production cross section with 
$t \approx t_{min} = -k_{z,min}^2 \approx 0$. 
We split  it  into the production cross section for 
low mass vector mesons $\rho,\,\omega$ and $\phi$, and 
the cross section for hadronic final states  carrying  
larger invariant masses $M_X^2 > 1\,GeV^2$:
\begin{equation} \label{eq:photonDD}
\left. \frac{d \sigma^{D}_{\gamma^{*} N}}{d M_X^2 dt} \right|_{t\approx 0} 
= 
\sum_{V=\rho,\omega,\phi}\left. 
\frac{d \sigma^{V}_{\gamma^{*} N}}{d M_X^2 dt} 
\right|_{t\approx 0} 
+ 
\left.\frac{d \sigma^{cont}_{\gamma^{*} N}}{d M_X^2 dt} 
\right|_{t\approx 0}. 
\end{equation}

First we explore this cross section 
in the kinematic domain of nuclear shadowing as measured 
by the E665 and NMC experiments. 
In this region at  $x\ll 0.1$ the experimentally accessible 
values for the momentum transfer  $Q^2$ are small. 
As an example, one has for $x<0.005$ an average momentum transfer 
$\overline {Q^2} <1\,GeV^2$ \cite{E665d,E665,NMC}. 
For a qualitative discussion, it is therefore 
useful first to consider diffractive (real) photoproduction processes 
for which  
more experimental data are available than in leptoproduction 
at finite $Q^2$.

\subsection{Photoproduction}

In the case of photoproduction the vector meson 
contribution to the diffractive cross section in (\ref{eq:photonDD}) 
can be described within the framework of 
generalized vector meson dominance \cite{Baueea78,NiPiWe96}. 
One finds: 
\begin{equation} \label{eq:resonanceDD}
\left. \frac{d \sigma^{V}_{\gamma N}}{d M_X^2 dt} \right|_{t\approx 0}
= \frac{e^2}{16\pi}\frac{\Pi^{V}(M_X^2)}{M_X^2} \sigma_{V N}^2,
\end{equation}
where 
\begin{equation}
\Pi(M_X^2) = \frac{1}{12 \pi^2} \frac{\sigma(e^+ e^- \rightarrow hadrons)}
{\sigma(e^+ e^- \rightarrow \mu^+\mu^-)}
\end{equation}
is the photon spectral function  as measured in $e^+e^-$ annihilation
and $e^2/4\pi = 1/137$. 
The contribution of the narrow $\omega$- and $\phi$-mesons  
to the photon spectral function is given by: 
\begin{equation}
\Pi^{V}(M_X^2) = 
\left(\frac{m_{V}}{g_{V}} \right)^2 
\delta(M_X^2 - m_{V}^2),
\end{equation}
for $V=\omega, \phi$, 
with the couplings $g_{\omega} = 17.0$, $g_{\phi } = 12.9$   
and masses  $m_{\omega} = 782\,MeV$, $m_{\phi}= 1019\,MeV$ 
\cite{Baueea78}. 
A comparison of (\ref{eq:resonanceDD}) with 
recent photoproduction data taken at HERA 
gives for the $\omega$--nucleon and $\phi$--nucleon cross sections 
at an average center of mass energy 
$\overline W = 80\,GeV$ and   $\overline W = 70\,GeV$, respectively:
\begin{equation}  
\sigma_{\omega N} = 26.0 \pm 2.5\,mb \,
\mbox{\cite{ZEUS96_om}} \quad \mbox{and}\;
\quad 
\sigma_{\phi N} = 19\pm 7\,mb\,  
\mbox{\cite{ZEUS96_ph}}.
\end{equation}
These values are in agreement with an  
analysis of nuclear effects in $\omega$- and $\phi$-production 
\cite{Baueea78} 
in the framework of vector meson dominance.  
Unlike  the $\omega$- and $\phi$-meson the $\rho$-meson has a 
large width from its strong coupling to two-pion states.
Therefore neither should the $\rho$-meson contribution to the 
photon spectral function be approximated by a $\delta$-function, 
nor is it  useful to separate contributions from resonant and non-resonant 
two-pion pairs. 
Both are accounted for in the $\pi^+\pi^-$ part of the photon 
spectral function which is given by the pion form factor $F_{\pi}$:  
\begin{equation} \label{eq:PiFpi}
\Pi^{\rho}\!\left(M_{X}^2\right) = \frac{1}{48 \pi^2} 
\Theta\!\left( M_{X}^2 - 4 m_{\pi}^2 \right) 
\left(1- \frac{4 m_{\pi}^2}{M_{X}^2}\right)^{3/2} 
\left| F_{\pi}\left(M_{X}^2\right)\right|^2,
\end{equation}
where $m_{\pi}$ and $M_X=M_{\pi\pi}$ are  the invariant masses  
of the pion and the  $\pi^+\pi^-$ pairs.
With an effective $\pi^+\pi^-$--nucleon cross section 
$\sigma_{\pi\pi N} = 30\,mb$ the main features of the mass 
distribution $d\sigma^{\pi\pi}_{\gamma N}/dM_{\pi\pi}$  
for $M_{\pi\pi}<1\,GeV$, 
as measured recently by the ZEUS collaboration \cite{ZEUS95_rho},  
are reproduced. 
Figure 1 shows our calculated  mass spectrum of $\pi^+\pi^-$ pairs 
using eqs.(\ref{eq:resonanceDD},\ref{eq:PiFpi}) in comparison with 
recent data from ZEUS \cite{ZEUS95_rho}. 
A perfect fit is obtained when small corrections from a possible 
mass dependence of $\sigma_{\pi\pi N}$ are included 
(for details see ref.\cite{NiPiWe96}).
For the pion form factor we have used the improved vector meson 
dominance form from ref.\cite{KlKaWe96} which reproduces the measured 
$F_{\pi}$ very well.

We note that the observed energy dependence of diffractive vector meson 
production is consistent with Regge phenomenology:  
\begin{equation}
\left.\frac{d\sigma^V_{\gamma N}}{dt}\right|_{t\approx 0}  
\approx  W^{4(\alpha_{\PP}(0) -1)} = W^{4\epsilon},
\end{equation}
where $\alpha_{\PP}(t=0) = 1+ \epsilon \approx 1.1$ is the soft pomeron 
intercept (see e.g. \cite{DonLan92}). 
This translates into an energy dependence of the effective 
vector meson-nucleon cross section:
\begin{equation} \label{eq:Pomeron} 
\sigma_{VN}\sim W^{2(\alpha_{\PP}(0) -1)} = W^{2\epsilon}
\approx W^{0.2}.
\end{equation}

The second term in eq.(\ref{eq:photonDD}) describes the  
$\sim 1/M_X^2$ behaviour of the diffractive production cross section
for large masses $M_X$. 
Guided by Regge phenomenology \cite{Goulia83} 
we use the parametrization:  
\begin{equation} \label{eq:PPP}
\left.\frac{d\sigma_{\gamma N}^{cont}}{d M_X^2 dt}\right|_{t\approx 0} 
= C \frac{W^{4 \epsilon}}{M_X^{2(1+\epsilon)}}.
\end{equation}
With $C = 8.2\,\mu b/GeV^{2(1+\epsilon)}$ we obtain a fair description 
of the measured diffractive cross section from ref.\cite{Chapin85} 
for $M_X>1.5\,GeV$ as shown in Fig.2.

\subsection{Leptoproduction}

Diffractive leptoproduction data have recently become available at  
large $Q^2$ ($8\, GeV^2 < Q^2 < 100 \, GeV^2$)  from the H1 
\cite{H195_diff,H196_diff} 
and ZEUS \cite{ZEUS95_diff,ZEUS96_diff,ICHEP_96} collaborations at HERA. 
The  corresponding diffractive production cross section 
is usually expressed in terms of 
the diffractive structure function $F_2^{D (4)}$. 
At small values $x<0.1$ of the Bjorken variable we have: 
\begin{equation} 
      F_2^{D (4)} (x,Q^2;x_{\PP}, t) = 
      \frac{Q^2}{\pi e^2} 
      \frac{d^2 \sigma^{D}_{\gamma^* N}}{d x_{\PP} dt}.
\end{equation}
Up to now no accurate data on  the $t$-dependence of 
$F_2^{D(4)}$  are available. 
Data exist only for the $t$-integrated structure function 
\begin{equation}
F_2^{D (3)} (x,Q^2;x_{\PP}) = \int_{-\infty}^{0} dt \, 
                            F_2^{D (4)} (x,Q^2;x_{\PP}, t). 
\end{equation}   
In the kinematic region of current HERA experiments 
one finds that $F_2^{D(3)}$ can be factorized as follows:  
\begin{equation} \label{eq:F(3)}
F_2^{D (3)} (x,Q^2;x_{\PP}) = \frac{A(\beta,Q^2)}{x_{\PP}^{n(\beta)}},
\end{equation} 
with $\beta = x/x_{\PP}$. 
Although first data on  $F_2^{D(3)}$ were consistent with 
a constant $n$ \cite{H195_diff,ZEUS95_diff,ZEUS96_diff}, most recent 
measurements from H1 indicate a 
weak $\beta$-dependence of  $n$ \cite{H196_diff}.

For the  investigation of nuclear shadowing 
we need the diffractive production 
cross section at $t\approx 0$, 
or equivalently $F_2^{D(4)}(x,Q^2;x_{\PP},t\approx 0)$, 
in a suitably parametrized form. 
We are guided again by Regge phenomenology 
and the picture, Fig.3, which suggests the following 
ansatz for $F_2^{D(4)}$
\cite{Regge_summ}
:
\begin{equation} \label{F(4)}
F_2^{D (4)} (x,Q^2;x_{\PP},t) = 
\frac{D(\beta,Q^2)}{x_{\PP}^{2 \alpha(\beta,t)-1}}
F^2(t).
\end{equation}
Here we have introduced the isoscalar nucleon form factor $F(t)$, as in 
high energy hadron-nucleon scattering \cite{DonLan88}. 
At small $t$ we use: 
\begin{equation}
F(t) = e^{b t/2},
\end{equation}
with $b \approx 4.6\,GeV^{-2}$.
We expand
\begin{equation}
\alpha(\beta,t) \approx \alpha_0(\beta) + \alpha_1(\beta) \,t + \dots
\end{equation}
For example, if $\alpha $ represents a soft pomeron we expect 
$\alpha_0 = \alpha_{\PP}(t=0) \approx 1.1$, 
and conventional Regge behavior corresponds to $\alpha_1 = 0.25\,GeV^{-2}$.
The remaining function $D(\beta,Q^2)$ needs to be determined by 
comparison with the available data on $F_2^{D(3)}$. 
We find: 
\begin{eqnarray} \label{eq:FD4}
F_2^{D(4)}(x,Q^2;x_{\PP},t=0) &=& 
\frac{D(\beta,Q^2)}{x_{\PP}^{2\alpha_0 - 1}}, \nonumber \\
&=& (b - 2 \alpha_1\, \ln \,x_{\PP}) \,F_2^{D(3)}(x,Q^2;x_{\PP}).
\end{eqnarray}
Omitting a possible logarithmic $Q^2$-dependence of the measured 
$F_2^{D(3)}$ we employ the parametrization \cite{ZEUS95_diff}: 
\begin{equation} \label{eq:F4A}
A(\beta,Q^2) = d\left[ \beta (1 - \beta) + \frac{f}{2} 
                       (1-\beta)^2\right]
\end{equation}
in eq.(\ref{eq:F(3)}). 
We then use  the following (different) sets of  
parametrizations motivated by the H1 and ZEUS data, respectively:
\begin{enumerate}
\item[(i)]  
We use $n = 1.19$ as found by the H1 collaboration 
\cite{H195_diff}.
Together with the conventional Regge value 
$0.25\,GeV^{-2}$ for $\alpha_1$ we take $d = 0.03$ and $f = 0.6$. 
The resulting $\alpha_0 \approx 1.13$ is close to 
the intercept $\alpha_{\PP}(0) \approx 1.1$ expected for 
a soft pomeron.

\item[(ii)]  The ZEUS collaboration recently found $n = 1.46$ 
\cite{ZEUS96_diff}, 
which seems to be incompatible with  the simple soft pomeron 
exchange picture\footnote{In previous \cite{ZEUS95_diff} 
and most recent \cite{ICHEP_96} ZEUS data $n\approx 1.3$ was found.
A reanalysis of the data is in progress \cite{Cart}.
}. 
In this case we choose 
$\alpha_1 = 0$, $d = 0.007$ and $f = 0.6$.
Here we get $\alpha_0 = (n+1)/2 = 1.23$. 
Note  that a ``hard'' pomeron  intercept 
$\alpha_0 \approx 1.5$ would lead to $n\approx 2$ which is too large.
 
\end{enumerate}

In Fig.4 we compare our parametrizations to the $x_{\PP}$-dependence 
of the measured structure function $F_2^{D(3)}$ 
as found by the H1 and ZEUS experiments. 
Note that a small fraction of the data includes diffractive 
dissociation of the nucleon target. In the present analysis we have 
not corrected for these events.
Furthermore we observe that the $t$-dependence, 
$F_2^{D(4)} \sim e^{B t}$, of 
the diffractive structure function is determined by the slope   
\begin{equation}
B(x_{\PP}) = b - 2 \alpha_1\,\ln \,x_{\PP}.
\end{equation}
In the kinematic range $10^{-4} < x_{\PP} < 0.05$ of HERA 
we find $B \approx (6$--$9)\,GeV^{-2}$ for the H1 set $(i)$ and  
$B \approx 5\,GeV^{-2}$ for the ZEUS set $(ii)$.

\section{Shadowing  effects in deep-inelastic scattering 
        \protect \\from deuterium}

The connection between diffractive (virtual) photoproduction 
and nuclear shadowing can be established most clearly for 
double scattering contributions to nuclear deep-inelastic scattering, 
i.e. processes in which the incoming photon beam interacts with two 
nucleons inside the  target\footnote{A similar analysis for 
hadron-nucleus collisions can be found in \cite{Had_Nucl}.}. 
To avoid complications from higher order multiple scattering  
we investigate shadowing effects for deuterium. 
This will be done in the laboratory frame.  
Realistic nuclear wave functions are established only in this frame.

The photon-deuteron forward scattering amplitude can be written 
as the sum of  single and double scattering contributions: 
\begin{equation}
{\cal A}_{\gamma^{*} d} = {\cal A}_{\gamma^{*} p}^{(1)} + 
                        {\cal A}_{\gamma^{*} n}^{(1)} + 
                        {\cal A}_{\gamma^{*} d}^{(2)}.
\end{equation} 
The ${\cal A}^{(1)}$ amplitudes describe the incoherent scattering of the  
(virtual) photon from the proton or neutron, 
while ${\cal A}^{(2)}$ accounts for the coherent interaction 
of the projectile with both nucleons. 
At large photon energies $\nu>3\,GeV$, or small values of $x < 0.1$ 
respectively, the double scattering amplitude  is dominated 
by diffractive excitations of the  photon to  hadronic 
intermediate states.
%Non-diffractive processes will hardly contribute:
%if a large momentum is transferred in the interaction with the 
%first nucleon, the overlap   
%with the wave function of the final state deuteron will be small.
%Furthermore, in rapidity space hadronic intermediate states  
%are well separated from the target nucleons, so that 
%final state interactions with the scattered nucleon can be neglected. 
We restrict ourselves to the dominant diffractive photoproduction 
in the forward direction described by the amplitude 
$T_{\gamma^{*} N \rightarrow XN}$.
Treating the deuteron target in the non-relativistic limit 
gives \cite{DS}:
\begin{equation} \label{eq:ds amplitude} 
{\cal A}_{\gamma^{*} d}^{(2)} = -\frac{1}{\pi M} 
\int_{-\infty}^{\infty} dz |\psi_d({\bf 0}_{\perp},z)|^2 
\sum_X\int_{-\infty}^{\infty} d k_z \,
T_{\gamma^{*} N\rightarrow XN}
\frac {e^{ik_z z}}{\nu^2 - (q_z + k_z)^2 - M_X^2 + i\epsilon } 
T_{XN\rightarrow \gamma^{*} N}. 
\end{equation}
Here $\psi_d(\vek r)$ is the deuteron wave function 
with the normalization $\int d\vek r |\psi_d(\vek r)|^2 = 1$.
We sum over all diffractively excited hadronic states $X$ 
with  invariant mass $M_X$ and four-momentum 
$(\nu,{\bf 0}_{\perp},q_z + k_z)$.
After integrating over the longitudinal momentum transfer 
$k_z$ the double scattering contribution 
to the total photon-deuteron cross section reads:
\begin{eqnarray} \label{eq:ds corr1}
\delta \sigma_{\gamma^{*}d} = 
\frac{Im {\cal A}_{\gamma^{*} d}^{(2)}}{4 M \nu} &=& 
- 16\pi \int_{4 m_{\pi}^2}^{W^2} dM_X^2 
\left. \frac{d^2\sigma^{D}_{\gamma^{*} N}}{dM_X^2 dt}
\right|_{t\approx 0}  
\,
\int_0^{\infty} dz \,|\psi_d({\bf 0}_{\perp},z)|^2 
\nonumber \\
&\times&
\left[ \left( 1 - 2 \left(\frac{Re \,T }
                             {Im \,T}\right)^2 \right) 
\cos(z/\lambda) - 
2 \frac{Re \,T}{Im \,T} \sin(z/\lambda) 
\right].  
\end{eqnarray}
Here $T\equiv T_{\gamma^{*} N\rightarrow XN}$, 
and $d^2\sigma^{D}_{\gamma^{*} N}/dM_X^2 dt|_{t \approx 0}$ 
is the forward diffractive cross section for the production of hadrons 
with invariant mass $M_X$:
\begin{equation}
\sum_X 
\left .\frac{|T_{\gamma^{*} N\rightarrow XN}|^2}
{(2 M \nu)^2} \right|_{t\approx 0}
= 16 \pi  \int_{4 m_{\pi}^2}^{W^2} dM_X^2 
\left. \frac{d^2\sigma^{D}_{\gamma^{*} N}}{dM_X^2 dt}
\right|_{t\approx 0}.
\end{equation}
In eq.(\ref{eq:ds corr1}) we sum over all diffractive excitations 
which are kinematically permitted, i.e. $4 m_{\pi}^2 \leq M_X^2 \leq W^2$. 
The longitudinal propagation length $\lambda$ of the intermediate hadronic 
state $X$ is the inverse of the minimal momentum 
transfer (\ref{eq:long})  necessary for its diffractive excitation:
\begin{equation}
\lambda = \frac{1}{k_{z,min}}=\frac{2\nu}{Q^2 + M_X^2}.
\end{equation}
Neglecting the real part of the diffractive production amplitudes leads 
to the well known result \cite{Gribov69}: 
\begin{eqnarray} \label{eq:ds corr2}
\delta \sigma_{\gamma^{*}d} = 
- 8\pi \int_{4 m_{\pi}^2}^{W^2} dM_X^2 
\left. \frac{d^2\sigma^{D}_{\gamma^{*} N}}{dM_X^2 dt}
\right|_{t\approx 0}  
{\cal F}_d(\lambda^{-1}), 
\end{eqnarray}
with the  longitudinal deuteron form factor 
\begin{equation} \label{eq:deuff}
{\cal F}_d(\lambda^{-1}) 
= 
\int_{-\infty}^{\infty} dz |\psi_d({\bf 0}_{\perp},z)|^2 
\cos(z/ \lambda).
\end{equation}
As soon as the longitudinal propagation length 
$\lambda(M_X^2)$ of  a certain hadronic intermediate state 
exceeds the average nucleon-nucleon distance  in 
the nuclear target, it will contribute to double scattering 
and therefore to shadowing. 
Since $\lambda$ decreases with $M_X$,  
low mass diffractive excitations are important 
for the onset of shadowing. 
If the propagation  length of a specific hadronic intermediate state 
exceeds the  target diameter,    
$\lambda > \langle r^2\rangle^{1/2} \sim 4\,fm$ for the deuteron, 
double scattering is not restricted since ${\cal F}_d \approx constant.$
Therefore in a first approximation hadronic states with an invariant mass 
\begin{equation} \label{eq:MX}
M_X^2 <M_{max}^2 =  \frac {W^2+ Q^2}{M \langle r^2\rangle^{1/2}} - Q^2
\end{equation} 
contribute to double scattering.  
Combining eqs.(\ref{eq:photonDD},\ref{eq:ds corr2},\ref{eq:MX}) 
yields the following approximate equation for the shadowing correction: 
\begin{eqnarray} \label{eq:approx}
\delta \sigma_{\gamma^{*} d} &\approx& 
- 8\pi {\cal F}_d(\lambda^{-1} \rightarrow 0)
\left\{
\int_{4 m_{\pi}^2}^{M_0^2} dM_X^2
\sum_{V=\rho,\omega,\phi}\left. 
\frac{d \sigma^{V}_{\gamma^{*}  N}}{d M_X^2 dt} 
\right|_{t\approx 0} + 
\int_{M_0^2}^{M_{max}^2} dM_X^2
\left.\frac{d \sigma^{cont}_{\gamma^{*}  N}}{d M_X^2 dt} 
\right|_{t\approx 0} 
\right\}, \nonumber\\
&=&  
- 8\pi {\cal F}_d(0) 
\left\{\sum_V b_V \,\sigma(\gamma^{*} N \rightarrow V N)  
+\, \overline {b_X} \,\sigma(\gamma^{*} N \rightarrow X N)\right\},  
\end{eqnarray}
where $M_0 \sim m_{\phi}$. 
Here we have assumed an exponential $t$-dependence, $e^{bt}$, for the 
diffractive production cross sections.  
The corresponding slopes for vector meson production   
are denoted by $b_V$, while $\overline{b_X}$  stands for the average slope 
for the production of hadronic states $X$ with 
$M_0^2 < M_X^2 < M_{max}^2$. 

Using  the  parametrizations for the diffractive cross sections 
as derived in Sec.2,   we  can now discuss shadowing 
in the deuteron for real and virtual photon  beams.

\subsection{Shadowing for real photons}

It is instructive to apply eq.(\ref{eq:approx}) to 
real photon-deuteron scattering. 
The relevant diffractive scattering cross sections 
for this case are given by eqs.(\ref{eq:resonanceDD},\ref{eq:PPP}). 
We find:  
\begin{eqnarray} \label{eq:approx photon}
\delta \sigma_{\gamma d}&\approx&  
- 8\pi {\cal F}_d(0) 
\left\{\sum_V b_V \,\sigma(\gamma N \rightarrow V N) + 
W^{4 \epsilon} \frac {C}{\epsilon} 
\left[ 
\frac{1}{M_0^{2 \epsilon}} - 
\left(\frac{M  \langle r^2\rangle^{1/2}}{W^2}\right)^{\epsilon} \right]
\right\}.  
\end{eqnarray}
We observe two sources for the energy dependence of the 
shadowing correction $\delta \sigma_{\gamma d}$.
First, the rising diffractive photoproduction cross sections 
$\sigma(\gamma  N \rightarrow X N) \sim W^{4 \epsilon}$ 
with $\epsilon \sim 0.1$ 
translate  into a contribution to shadowing with a similar 
energy dependence.
However this is not the only source for an increase of 
$\delta\sigma_{\gamma d}$ with $W$. 
With rising center of mass energy  the propagation 
length $\lambda$ also increases and allows for additional contributions to 
shadowing from diffractively produced states of 
large mass as can be seen from eq.(\ref{eq:MX}). 

In Fig.5 we show the shadowing correction $\delta \sigma_{\gamma d}$ 
calculated from eq.(\ref{eq:ds corr2}). 
Here a realistic deuteron form factor (\ref{eq:deuff}) as obtained 
from the Paris potential \cite{LaLoRi80} was used.
For $\nu > 50\,GeV$ we find good agreement with the approximation 
in eq.(\ref{eq:approx photon}). 
The vector meson contribution to shadowing is proportional 
to $\nu^{2\epsilon}\approx \nu^{0.2}$ and dominates 
$\delta \sigma_{\gamma N}$ for $\nu <300\,GeV$. 
At higher energies contributions from hadronic 
states of large mass also become important.  
In the energy range of present fixed target experiments, $\nu < 400\,GeV$, 
their increase with $\nu$ is stronger than $\nu^{0.2}$ due to the 
additional contributions from large mass states.

As already mentioned, recent data on nuclear shadowing from 
the E665 and NMC collaborations were taken at small 
momentum transfers $\overline {Q^2}\lsim 1\,GeV^2 $ for $x\ll 0.1$. 
The energy transfer in these experiments is typically 
$40\,GeV < \nu < 380\,GeV$ \cite{E665d,E665,NMC}. 
Therefore the conclusion just drawn applies here too: 
nuclear shadowing as measured by  
E665 and NMC is dominated by the diffractive 
excitation and multiple scattering of vector mesons. 
This observation is in agreement with a number of model 
calculations (see e.g. \cite{BadKwi92npb,MelTho93,PiRaWe95}). 
Nevertheless large mass states play their role even 
in the kinematic region of the E665 and NMC measurements 
as they are responsible for the leading twist nature 
of nuclear shadowing (see e.g. \cite{PiRaWe95,FraStr89,NikZak91zpc} and 
references therein). 

In Fig.6 we show the ratio of the total photon-deuteron cross section 
compared to twice the free photon-nucleon cross section, 
$R = \sigma_{\gamma d}/2\sigma_{\gamma N}$.
For the free photon-nucleon cross section we use the empirical 
photon-proton cross section from \cite{FNAL78} with the energy 
dependence $\sigma_{\gamma N} \sim W^{2 \epsilon}\approx W^{0.2}$. 
Then the  shadowing correction grows as 
$\delta \sigma_{\gamma d}/2 \sigma_{\gamma d} = R-1 \sim 
\nu^{\epsilon}\approx \nu^{0.1}$. 
We find that our results are well within the range of recent lepton-deuteron 
scattering data from the E665 collaboration \cite{E665d} 
measured at $\overline {Q^2} \lsim 0.5\,GeV^2$.

\subsection{Shadowing at large $Q^2$} 

When expressed in terms of the diffractive structure function 
the double scattering contribution  to 
the deuteron structure function, 
$\delta F_{2d} = \frac{Q^2}{\pi e^2} \,\delta \sigma_{\gamma^* d}$,
becomes (\ref{eq:approx}): 
\begin{equation} \label{eq:approx_F} 
\delta F_{2d}(x,Q^2) \approx - 8 \pi {\cal F}_d(0) 
\int_{x_0}^{x_m} d x_{\PP} \,
F_2^{D(4)}(x,Q^2;x_{\PP},t\approx 0),
\end{equation}
where $x_m \approx 1/M\langle r^2 \rangle ^{1/2}$ 
and $x_0 \approx Q^2/W^2\approx x$ for $x\ll 0.1$. 
An investigation of $\delta F_{2d}$ using the parametrizations 
for $F_2^{D(4)}$ as derived in Sec.2.2 yields, that shadowing 
is dominated at $x\ll 0.01$ by the leading contribution to the diffractive 
structure function in the limit of small $x_{\PP} \ll 0.01$. 
Using  eq.(\ref{eq:FD4}) we obtain at small $x \ll 0.1$: 
\begin{equation} \label{eq:delF2d_approx}
\delta F_{2d}(x,Q^2) \approx - 8 \pi {\cal F}_d(0) 
\,
\frac{1}{x^{n - 1}} 
\frac{d}{n}
\left\{
\left(b - 2 \alpha_1\,\ln\,x \right) \frac{f+n-1}{n^2 - 1}
- 2 \alpha_1 \frac{(1-f)(1-3n^2) + 2n^3}{n(n^2 -1)^2}
\right\}.
\end{equation}
For the different parametrizations of the diffractive 
structure function $F_2^{D(4)}$ in Sec.2.2 
we find from eq.(\ref{eq:delF2d_approx})  
that at $x\ll 0.01$ shadowing increases with 
decreasing $x$ as $\delta F_{2d} \sim x^{-0.25}$ for 
the parameter set $(i)$, and $\delta F_{2d} \sim x^{-0.46}$ for 
$(ii)$. 
This observation is confirmed by an explicit calculation of 
the double scattering correction using eq.(\ref{eq:ds corr2}) 
together with the  
deuteron form factor as obtained from the Paris potential.
The result for $\delta F_{2d}$ is shown in Fig. 7. 
For $x<0.01$ it agrees well with the 
approximation in (\ref{eq:delF2d_approx}). 
We conclude that at large $Q^2 \gsim 10\,GeV^2$ and small 
$x < 0.01$ shadowing is controlled 
by the behavior of the diffractive structure function at 
small values of $x_{\PP}$, or equivalently at large hadronic masses 
$M_X$. 

Recent data from HERA show that at corresponding values of  
$x\ll 0.01$ and $Q^2 \gsim 10\,GeV^2$ the inclusive  nucleon structure 
function  rises as $F_{2N} \sim x^{-\lambda }$ with $\lambda = 0.25-0.3$ 
\cite{HERA_F2N}. 
Therefore we find that in this kinematic region 
$R = F_{2d}/2 F_{2N}$ increases moderately with decreasing $x$  for 
the parameter set $(i)$  which is guided by the H1 data on 
the diffractive structure function. 
For the parameter set $(ii)$ motivated by data from ZEUS  
we observe a slow decrease of the shadowing ratio with decreasing $x$. 
In Fig.8 we present the results for the structure function 
ratio $R$ using the free nucleon structure function  
from ref.\cite{CTEQ94} taken at $Q^2=25\,GeV^2$. 

A comparison with the results 
from the previous section demonstrates  the qualitative difference 
between  the energy dependence of shadowing at small and large $Q^2$.

A similar prediction was made recently by Kopeliovich and Povh 
\cite{KopPov96}.
They express the virtual photon-nucleon cross section at small $x$ 
through  the interaction of hadronic fluctuations of the virtual photon 
with different transverse sizes.
The contribution of a certain hadronic fluctuation 
is given by its weight in the photon wave function, times its 
interaction cross section.
The probability to find large size ($\sim 1/\Lambda_{QCD}$) quark-gluon 
configurations in the virtual photon is suppressed by 
$\Lambda_{QCD}^2/Q^2$,  
as compared to small configurations with transverse sizes 
$b^2 \sim Q^{-2}$. 
However since the 
interaction cross sections of such hadronic fluctuations are proportional 
to their transverse size, 
both give leading contributions $\sim 1/Q^2$  
to the photon-nucleon cross section.
This is different for the coherent interaction of the virtual 
photon with several nucleons as, in deep-inelastic scattering from 
nuclei at small $x$. 
For example the contribution of a hadronic fluctuation to  
double scattering is given by its weight in 
the photon wave  function, multiplied by  the {\it square} of 
its interaction cross section. 
Consequently large size configurations yield a leading $1/Q^2$--contribution 
to the double scattering cross section, 
while contributions from small size configurations are proportional 
to $1/Q^4$ and suppressed.
Therefore multiple scattering contributions to the virtual photon-nucleus 
cross section at small $x$ are dominated by the interaction of 
large-size hadronic fluctuations.
The energy dependence of the interaction cross section of 
photon-induced large size 
quark-gluon configurations should be similar to the one observed in 
high energy hadron-hadron collisions, and 
therefore  much weaker than the  strong 
energy dependence of the nucleon structure function at small $x$ 
and large $Q^2$.

Finally we emphasize that in contrast to the  $Q^2\sim 0$ 
case discussed in Sec.3.1,  
the diffractive excitation of vector mesons is  not relevant  
at large $Q^2\gsim 10\,GeV^2$. 
This is simply a consequence of the strong decrease 
of vector meson production cross sections with $Q^2$,
e.g. $\sigma(\gamma^* N \rightarrow \rho N) \sim Q^{-\beta}$
with $\beta \approx 4.2$--$5.0$ \cite{VMQ2ALL}. 
From eq.({\ref{eq:approx}) we estimate the vector meson contribution 
to the shadowing correction $\delta F_{2d}$ to be less than 
$3\%$ at large $Q^2$.

\section{Summary}
Shadowing effects in deep-inelastic lepton-nucleus 
scattering are dominated by the diffractive excitation 
and coherent interaction of hadronic Fock states 
present in the wave function of the exchanged 
virtual photon. 
Therefore shadowing is closely related to diffractive 
leptoproduction of hadrons from individual nucleons. 
We have investigated this connection using 
the currently available 
experimental information  for both processes.
We find that in the kinematic regime of  
small $x\ll 0.1$ and small $Q^2 < 1\,GeV^2$, as in 
recent experiments from the NMC and E665 collaboration,
shadowing is dominated by vector mesons.
At large values of $Q^2$ the situation is different. 
Using data on the diffractive nucleon structure function 
from HERA we have shown that heavy mass states play a dominant role in  
shadowing at small  $x \ll 0.1$ and large $Q^2\gg 1\,GeV^2$. 
Furthermore we have demonstrated that the energy dependence 
of shadowing is directly related to the dependence of 
the diffractive production cross section on the mass 
of the final hadronic state. 
In view of these results, it appears that 
the currently discussed option of accelerating ions 
at HERA could add interesting new information about diffractive 
processes for nuclear and nucleon targets.

\newpage
{\centerline {\bf Figure Captions} }

\begin{enumerate}

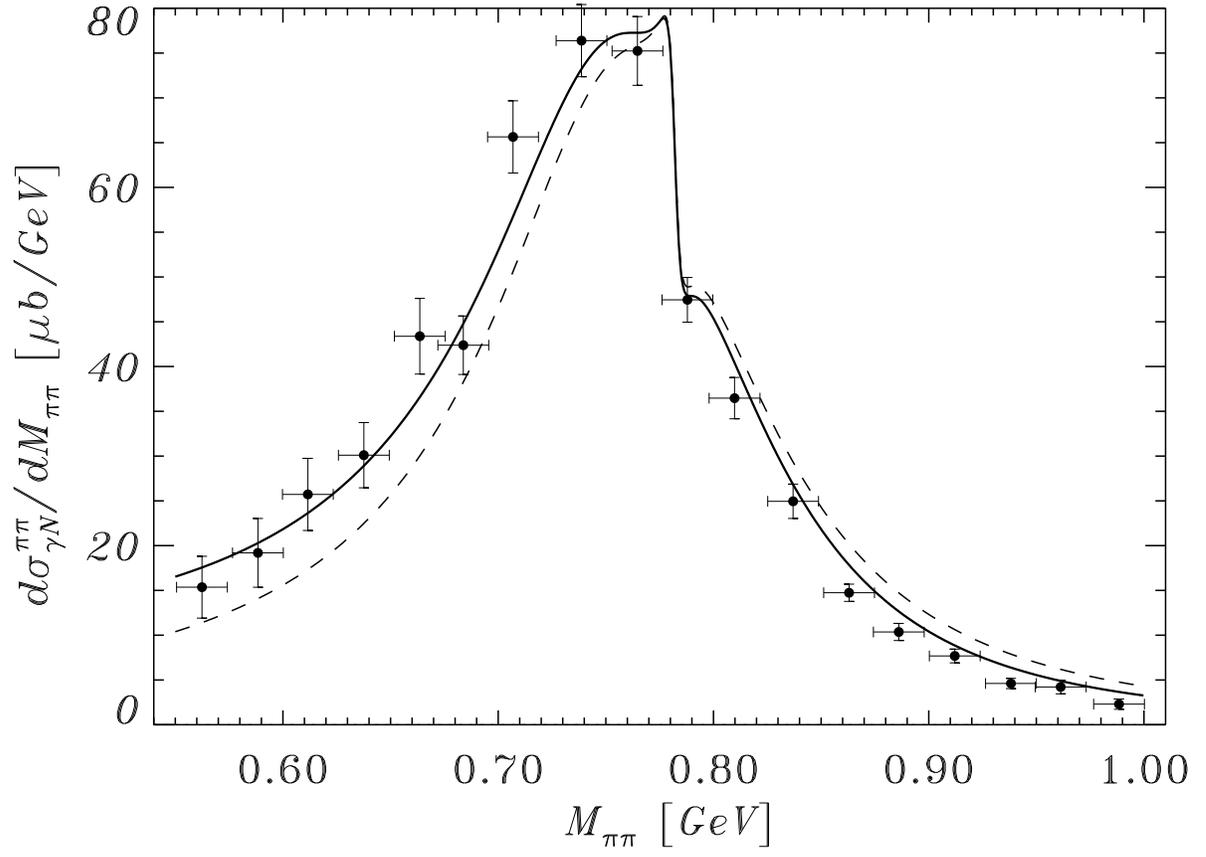
\item[Figure 1:] 
The mass distribution $d\sigma_{\gamma N}^{\pi\pi}/dM_{\pi\pi}$ of 
diffractive 
photoproduction in the rho meson region. The dashed curve 
shows the contribution from the 
$\pi^+ \pi^-$ part of the photon spectral function (\ref{eq:PiFpi}). 
The full curve includes a small mass dependence of 
$\sigma_{\pi\pi N}$ \cite{NiPiWe96}. 
The data are taken from the ZEUS collaboration \cite{ZEUS95_rho}.

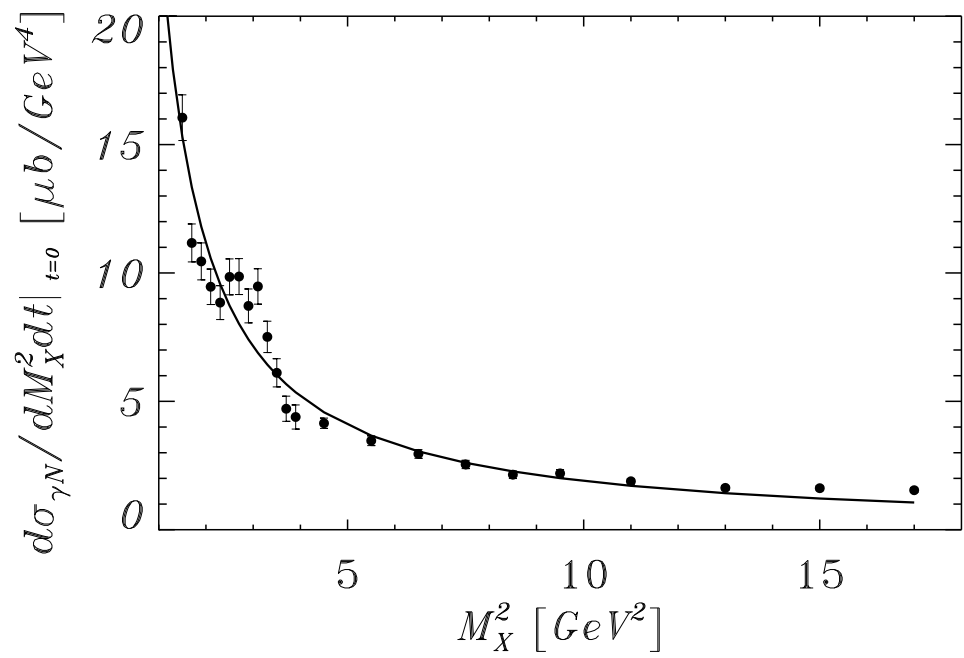
\item[Figure 2:]
The diffractive photoproduction cross section for large 
masses $M_X$. The full curve shows the fit from eq.(\ref{eq:PPP}).
The data are taken from \cite{Chapin85} and extrapolated 
to $t=0$ using an average slope $b = 5\,GeV^{-2}$.

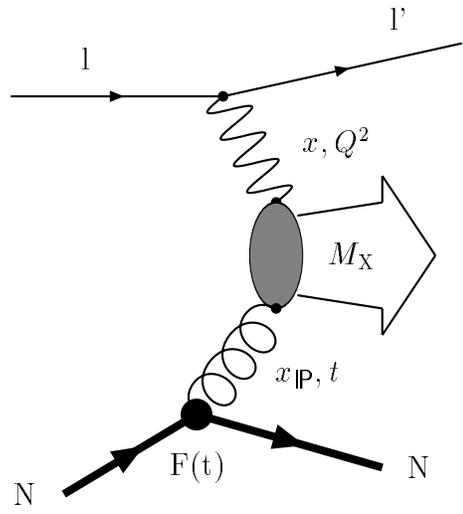
\item[Figure 3:]
Regge picture of diffractive lepton-nucleon scattering \cite{Regge_summ}. 

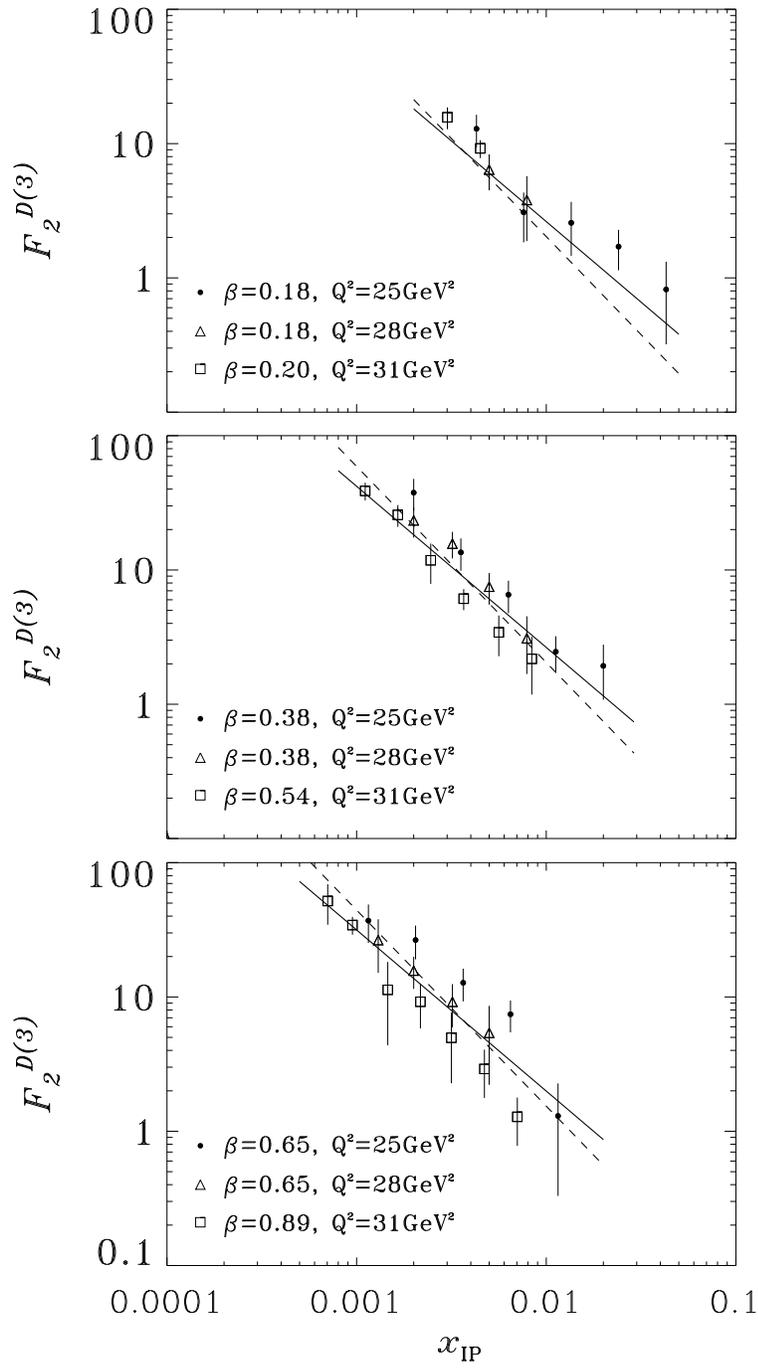
\item[Figure 4:]
The diffractive structure function $F_2^{D(3)}$ 
as a function of $x_{\PP}$. The full and dashed curves correspond to 
the parametrizations $(i)$ and $(ii)$ in Sec.2.2.
The data are taken from  H1 (dotted) \cite{H195_diff} and  
ZEUS (triangles) \cite{ZEUS95_diff},
(squares) \cite{ZEUS96_diff}. 

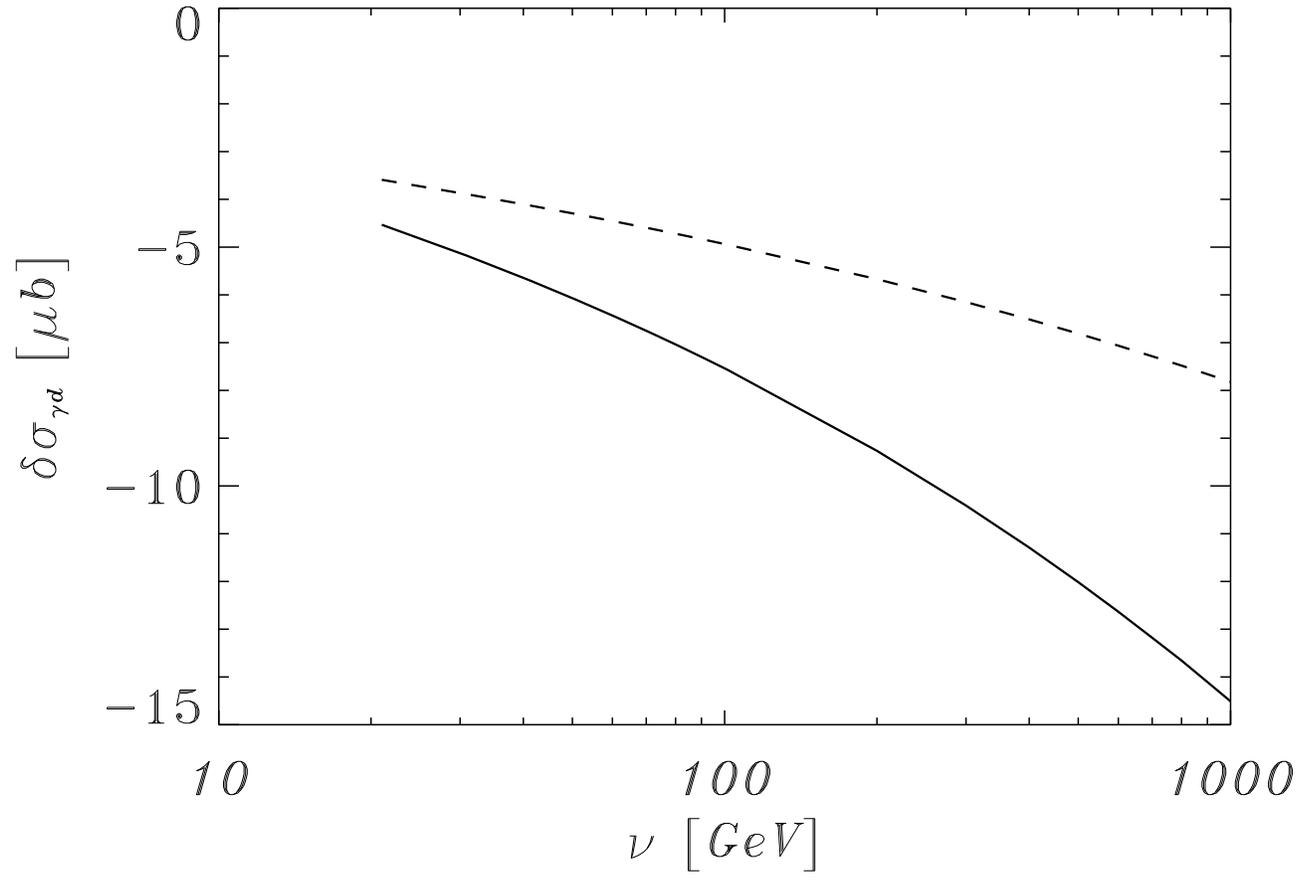
\item[Figure 5:]
The energy-dependence of the shadowing correction 
$\delta \sigma_{\gamma d}$ calculated from eq.(\ref{eq:ds corr2}). 
The dashed line shows the vector meson contribution. 
 
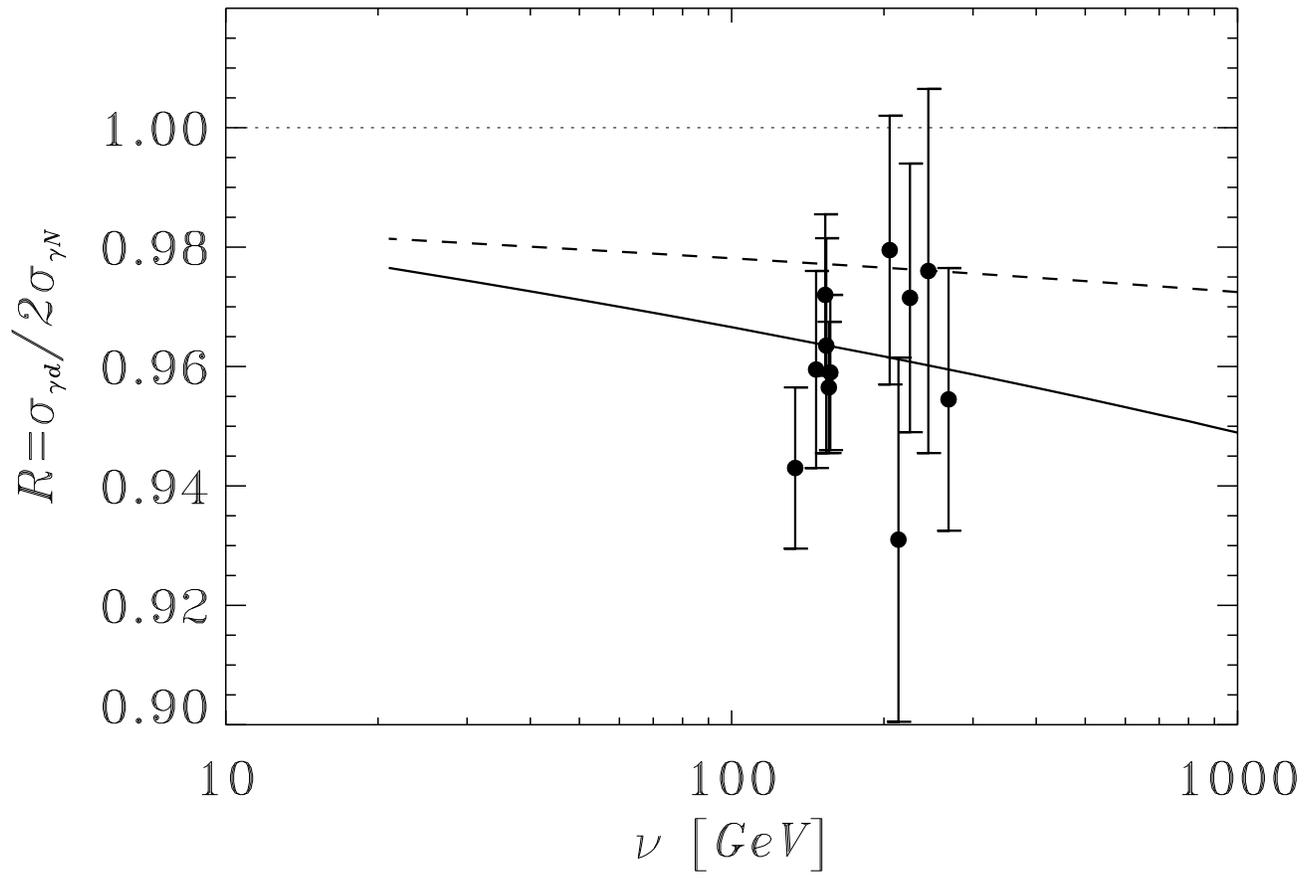
\item[Figure 6:]
The shadowing ratio $ R = \sigma_{\gamma d}/2 \sigma_{\gamma N}$ 
as a function of the photon energy. 
The dashed line shows the vector meson contribution. 
The experimental data are taken from the E665 collaboration 
\cite{E665d}. (The energy values of the data have to be understood 
as average values which correspond to different $x$-bins.)

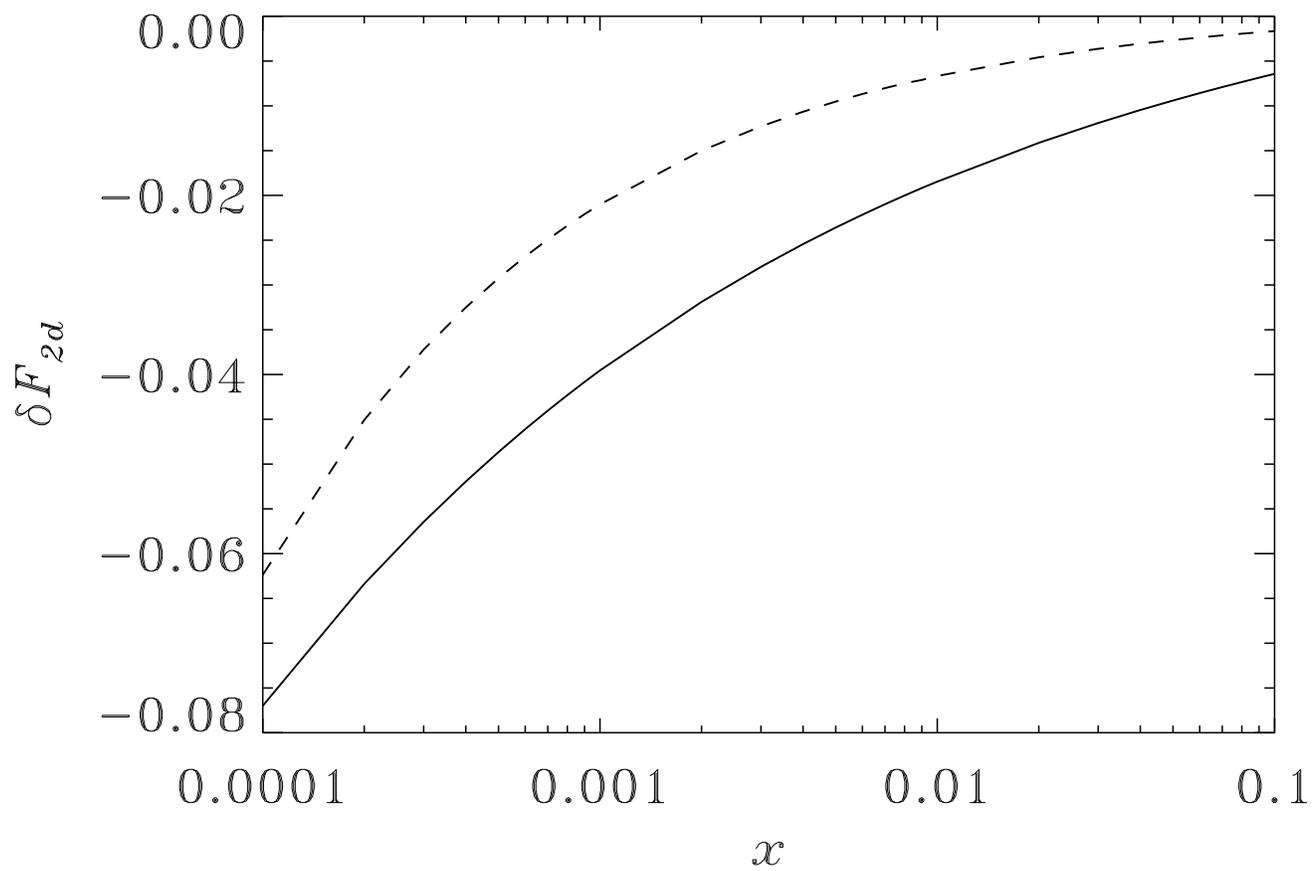
\item[Figure 7:]
The shadowing correction $\delta F_{2d}$ 
from eq.(\ref{eq:ds corr2}) at large 
$\overline {Q^2} \sim 25\,GeV^2$ plotted against $x$.
For the diffractive structure function $F_2^{D(4)}$ 
we have used the parametrizations $(i)$ (full) and $(ii)$ (dashed) 
from Sec.2.2.

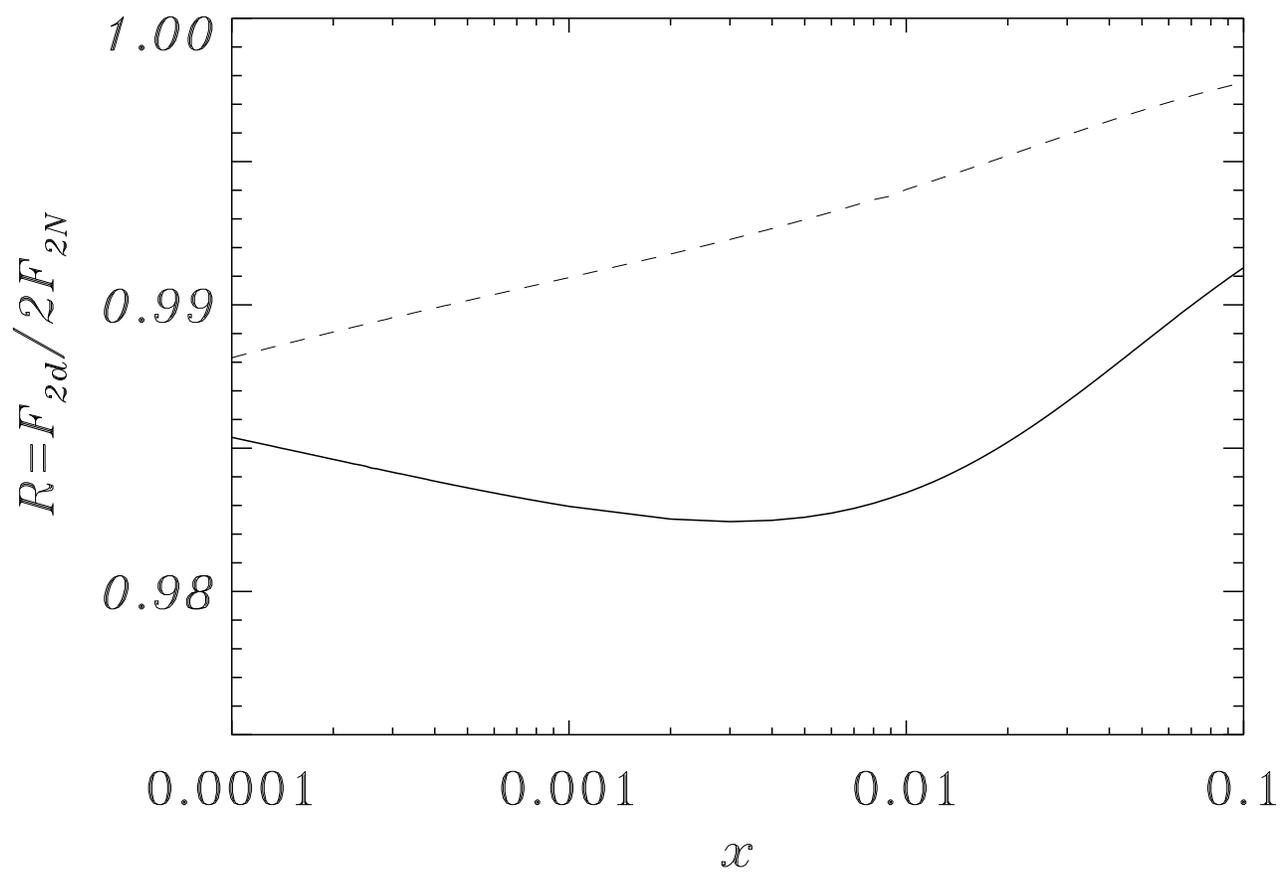
\item[Figure 8:]
The shadowing ratio $R=F_{2d}/2 F_{2N}$ at large 
$\overline {Q^2} \sim 25\,GeV^2$. 
The full and dashed curve correspond to the diffractive 
structure function  $(i)$  and $(ii)$  from Sec.2.2, 
respectively. 
The nucleon structure function has been taken from 
\cite{CTEQ94} at $Q^2 = 25\,GeV^2$.

\end{enumerate}

\end{document}